\documentclass[%
reprint,
superscriptaddress,
 amsmath,amssymb,
 aps
]{revtex4-1}

\usepackage[english]{babel}
\usepackage{amsmath}
\usepackage{amssymb}
\usepackage{braket}
\usepackage{graphicx}
\usepackage{soul}
\usepackage{multirow}
\usepackage{commath}
\usepackage{stmaryrd}
\usepackage{booktabs}
\usepackage{threeparttable}
\usepackage{comment}
\usepackage[percent]{overpic}
\usepackage{tikz}
\usetikzlibrary{positioning}
\usepackage{xr-hyper}
\usepackage[colorlinks=true,urlcolor=blue,citecolor=blue,linkcolor=blue]{hyperref}
\usepackage[normalem]{ulem}
\usepackage[a4paper,top=3cm,bottom=3cm,left=2.5cm,right=2.5cm]{geometry}
\usepackage{footnote}

\begin{document}

\title{Local control theory for superconducting qubits}

\author{M. Mali\v{s}}
\affiliation{Centre Europ\'een de Calcul Atomique et Mol\'eculaire, Ecole Polytechnique F\'ed\'erale de Lausanne, Avenue Forel 2, 1015 Lausanne, Switzerland}

\author{P. Kl. Barkoutsos}
\affiliation{IBM Research GmbH, Zurich Research Laboratory, S\"aumerstrasse 4, 8803 R\"uschlikon, Switzerland}
\affiliation{Institute for Theoretical Physics, ETH Zurich, 8093 Zurich, Switzerland}
\author{M. Ganzhorn}
\affiliation{IBM Research GmbH, Zurich Research Laboratory, S\"aumerstrasse 4, 8803 R\"uschlikon, Switzerland}
\author{S. Filipp}
\affiliation{IBM Research GmbH, Zurich Research Laboratory, S\"aumerstrasse 4, 8803 R\"uschlikon, Switzerland}
\author{D. J. Egger}
\affiliation{IBM Research GmbH, Zurich Research Laboratory, S\"aumerstrasse 4, 8803 R\"uschlikon, Switzerland}
\author{S. Bonella}
\affiliation{Centre Europ\'een de Calcul Atomique et Mol\'eculaire, Ecole Polytechnique F\'ed\'erale de Lausanne, Avenue Forel 2, 1015 Lausanne, Switzerland}
\author{I. Tavernelli}
\email{ita@zurich.ibm.com}
\affiliation{IBM Research GmbH, Zurich Research Laboratory, S\"aumerstrasse 4, 8803 R\"uschlikon, Switzerland}

\date{\today}

\begin{abstract}
In this work, we develop a method to design control pulses for fixed-frequency superconducting qubits coupled via tunable couplers based on local control theory, an approach commonly employed to steer chemical reactions. 
Local control theory provides an algorithm for the monotonic population transfer from a selected initial state to a desired final state of a quantum system through the on-the-fly shaping of an external pulse. 
The method, which only requires a unique forward time-propagation of the system wavefunction, can serve as starting point for additional refinements that lead to new pulses with improved properties. 
Among others, we propose an algorithm for the design of pulses that can transfer population in a reversible manner between given initial and final states
of coupled fixed-frequency superconducting qubits.

\end{abstract}

\maketitle

\section{Introduction}
\label{sec: Intro}

Methods for pulse shaping~\cite{Glaser2015} to control quantum processes have allowed important advances in different domains, ranging from the steering of photo-chemical processes~\cite{shapiro2003principles,Balint2008,Sola2018} to the optimization of gate operations in quantum computing~\cite{Khaneja2005_grape,Motzoi2009}.

In quantum information, optimal control theory (OCT) is typically used to generate target unitary operators~ \cite{Khaneja2005_grape, Caneva2011_crabe, Kelly2014, Egger2014, Machnes2018_goat}.
Within the field of superconducting qubits~\cite{Devoret2013} OCT has been successfully applied to design various qubit gates in different hardware implementations \cite{Motzoi2009, Egger2013a, Schutjens2013_wahwah, Liebermann2017, Heeres2017} as well as to identify optimal operating conditions, such as the quasi-dispersive regime~\cite{Goerz2017}.

In parallel to OCT, local control theory (LCT) has also emerged as a valuable approach to control the dynamics of quantum systems by shaping external fields.
In particular, LCT has already been successfully applied to steer photo-chemical reactions in molecular systems~\cite{Marquetand2007, Volker2009, Curchod2015}. 
In LCT, an external field is designed on-the-fly under the constraint that it monotonically increases the quantum population of a selected target state when starting from a given initial state~ \cite{Kosloff1992, Curchod2011}.
While OCT is based on a computationally intensive variational approach, which requires computing the full time evolution of the system at each optimization step, LCT can generate pulses that produce the desired population transfer by computing the evolution of the system only once.
Although LCT does not necessarily provide a time-optimal pulse, thanks to its remarkable computational efficiency and conceptual simplicity, it can nonetheless become the method of choice for the design of state preparation pulses.

In this paper, we focus our investigation on the application of LCT to generate state preparation pulses for  fixed-frequency superconducting qubits coupled via tunable couplers. 
In Sec.~\textcolor{red}{\ref{sec: methods}}  we introduce LCT and show how to apply it to a setup made-up of fixed-frequency transmon qubits coupled by a tunable coupler~\cite{McKay2016, Roth2017}. 
Sec.~\ref{subsec:bare_lct} presents and discusses the pulses generated by the LCT algorithm. 
Sec.~\ref{subsec:lct_optimization}-\ref{subsec:truncation} reports on a procedure aimed at further optimizing their properties such as bandwidth, pulse length, and gate reversibility.  

\section{Methods}
\label{sec: methods}

\subsection{Theoretical background}
\label{subsec: theory}

We consider $n$ fixed-frequency qubits all mutually interacting through a single flux-tunable qubit, called tunable coupler (TC)~\cite{McKay2016}. 
Such systems combine the long coherence time of fixed-frequency transmon qubits with the high controllability of flux-tunable coupling elements.
The system is described by the Hamiltonian~\cite{Roth2017}
\begin{align}
\hat{H}(t) =& -\frac{1}{2} \sum_{i=1}^{n} \omega_{i} \hat{\sigma}_{i}^{z} + \sum_{i=1}^{n} g_{i} \left( \hat{\sigma}_{i}^{+} \hat{\sigma}_{\text{TC}}^{-} +\hat{\sigma}_{i}^{-} \hat{\sigma}_{\text{TC}}^{+} \right)  \nonumber \\
&-\frac{1}{2} \omega_{\text{TC}}(t)  \hat{\sigma}_{\text{TC}}^{z} \text{ ,} \label{eq1}
\end{align}
in units of $\hbar=1$. 
The qubit $i$ and TC raising and lowering operators are $\hat\sigma_i^+$, $\hat\sigma_i^-$, $\hat\sigma_\text{TC}^+$ and $\hat\sigma_\text{TC}^-$, respectively, while the number operators are $\hat\sigma^z_i$ and $\hat\sigma^z_\text{TC}$.
Qubit $i$ has frequency $\omega_i$ and couples with strength $g_i$ to the TC.
The frequency of the TC, $\omega_\text{TC}(t)$, is controlled by a current $I(t)$ brought close to the TC by a high-speed flux bias line, see Fig.~\ref{fig1}(a). 
The resulting flux $\Phi(t)$ threading through the TC SQUID loop changes the frequency of the TC according to 
\begin{align} \label{Eqn:TCPhi}
\omega_{\text{TC}}(t) = \omega^{0}_{\text{TC}} \sqrt{ | \cos{(\pi \Phi(t) / \Phi_{0})} | } \text{,}
\end{align}
where $\Phi_{0}$ is the magnetic flux quantum \cite{Koch2007}. 
The full system wave function $\ket{\Psi(t)}$ then evolves according to the time-dependent Schr\"{o}dinger equation 
\begin{equation}
\imath \partial_t \ket{\Psi(t)} = \hat{H}(t)\ket{ \Psi(t) }\text{.} 
\label{eq6}
\end{equation}
The population $\langle\hat{P}_\phi\rangle$ of any $n$-qubit target state $\ket{\phi}$ is governed by
\begin{equation}
\partial_t \langle \hat{P}_{\phi} \rangle = \imath \langle \left[ \hat{H}(t), \hat{P}_{\phi} \right] \rangle\text{,} \label{eq5}
\end{equation}
where $\hat{P}_{\phi} = \ket{ \phi }\!\!\bra{ \phi }$ is the corresponding projector operator and $\langle \dots \rangle$ denotes the expectation value with respect to $\ket{ \Psi(t)}$.
In our model, the only free, tunable parameter is the frequency of the tunable coupler $\omega_{\text{TC}}(t)$. 
We will, thus, employ LCT to increase the population in $\ket{\phi}$ by shaping $\omega_\text{TC}(t)$ on-the-fly. 
The TC frequency can be decomposed into a time-independent and a time-dependent part $\omega_{\text{TC}}(t) = \omega_{\text{TC}} + \delta\omega_{\text{TC}}(t)$ \cite{Curchod2011, Curchod2015}.
This splits the Hamiltonian $\hat{H}(t)$ into a time-dependent $\hat{H}^{\prime}(t)=-\delta\omega_\text{TC}(t) \hat\sigma_\text{TC}^{z}/2$ and a drift term
\begin{align}
\hat{H}_{\text{d}} =& -\frac{1}{2} \sum_{i=1}^{n} \omega_{i} \hat{\sigma}_{i}^{z} + \sum_{i=1}^{n} g_{i} \left( \hat{\sigma}_{i}^{+} \hat{\sigma}_{\text{TC}}^{-} +\hat{\sigma}_{i}^{-} \hat{\sigma}_{\text{TC}}^{+} \right) \nonumber \\
& -\frac{1}{2} \omega_{\text{TC}}  \hat{\sigma}_{\text{TC}}^{z}\text{.} \label{eq3}
\end{align}
The drift $\omega_{\text{TC}}$ term depends on the constant DC flux  bias applied to the TC~\cite{McKay2016}.
When the target state $\ket{\phi}$ is an eigenvector $\ket{\psi_{j}}$ of the drift Hamiltonian $\hat{H}_{\text{d}}$ the projector operator $\hat{P}_{\phi}$ commutes with $\hat{H}_{\text{d}}$
and Eq.~\eqref{eq5} simplifies to
\begin{equation}
\partial_t \langle \hat{P}_{j} \rangle = -\frac{\imath}{2} \delta\omega_{\text{TC}}(t) \langle \left[ \hat{\sigma}_{\text{TC}}^{z}, \hat{P}_{j} \right] \rangle \text{.} \label{eq7}
\end{equation}
LCT induces a monotonous increase of the target state population by generating a $\delta\omega_{\text{TC}}(t)$ pulse that guarantees the positivity of the right hand side of Eq.~\eqref{eq7}. 
For our setup this condition is achieved by changing the frequency of the TC according to
\begin{equation}
\delta \omega_{\text{TC}}(t) = \frac{\imath}{2} \lambda 
\langle \left[ \hat{\sigma}_{\text{TC}}^{z}, \hat{P}_{j} \right] \rangle^* 
\text{.} \label{eq8}
\end{equation}
The coupling parameter $\lambda$ controls the magnitude by which the control field $\delta \omega_{\text{TC}}(t)$ is changed.
Its value can be tuned as long as the resulting pulse $\delta \omega_{\text{TC}}(t)$ can be implemented in realistic experimental setups. 

In the case of large systems with many possible states, the implementation of the LCT scheme can become numerically challenging. 
However, when some of the (high energy) states do not contribute to the dynamics, we can restrict the action of the LCT algorithm to a subspace of the full Hilbert space using the projector operator $\hat{P}_{n'}=\sum_{k=1}^{n'} \ket{ \psi_{k} }\!\bra{ \psi_{k} }$ over the first  $n'$ eigenvectors (assumed to be ordered according to their corresponding eigenvalues). Equation (\ref{eq8}) then simplifies to
\begin{align} \label{eq_proj}
\delta \omega_{\text{TC}}(t) \simeq \\
- \lambda \text{ Im} \Big( \sum_{k}^{n'} & \bra{ \psi_{j} } \hat{\sigma}_{\text{TC}}^{z} \ket{ \psi_{k} } \braket{ \psi_{k} | \Psi(t) }  
\braket{ \psi_{j} | \Psi(t) }^* \notag 
\Big) \text{.}
\end{align}
Since the TC frequency cannot exceed $\omega^0_\text{TC}$, see Eq.~\eqref{Eqn:TCPhi}, $\delta\omega_{\text{TC}}(t)$ is confined to the interval $[-\omega^0_{\text{TC}},0]$.
Thus, it is  necessary to impose a restriction on the $\lambda$-factor in order to avoid reaching the upper bound of $\delta\omega_{\text{TC}}(t)$. 
This is accomplished by capping the value of $\delta\omega_{\text{TC}}(t)$ to 0 (i.e. taking $\min[\delta\omega_{\text{TC}}(t),0]$) and by  
constraining the magnitude of $\lambda$ such as $\delta\omega_{\text{TC}}(t)>-\omega_{\text{TC}}^{0}$.

The LCT algorithm can be summarized in two steps: First, the instantaneous state 
is propagated for a short time interval $[ t, t + \delta t]$ under $\hat H(t)$. 
Second, the resulting wavefunction $\ket{ \Psi(t+\delta t)}$ is used to update the external field using Eq.~\eqref{eq_proj}. 
These two steps are repeated using the updated control field until the desired population transfer is achieved.
A smooth external driving pulse is obtained when $\delta t$ is made sufficiently small. 
For practical purposes, when the target state $\ket{ \phi }$ does not overlap with the initial system wavefunction $\ket{ \Psi(0) }$ a small fraction $\eta$ of the target state is added into the initial wavefunction 
\begin{align} \label{Eqn:seed}
\ket{ \Psi^{\prime}(0) } = \sqrt{\eta} \ket{ \psi_{j} } + \sqrt{1-\eta} \ket{ \Psi(0) }
\end{align}
to ensure that the LCT algorithm converges.

\subsection{System}
\label{subsec: system}
We apply LCT to a system composed of $n=2$ qubits, see Eq.~\eqref{eq1} and Fig.~\ref{fig1}(a). 
The qubits, labeled Q1 and Q2, are set at realistic~\cite{McKay2016} frequency values, $\omega_{1}/(2 \pi)=5.890~\rm{GHz}$ and $\omega_{2} / (2 \pi)=5.031~\rm{GHz}$, respectively. 
They are coupled with strengths $g_{1}/(2 \pi)=100~\rm{MHz}$ and $g_{2} / (2 \pi)=71~\rm{MHz}$ to a TC with a  maximal frequency $\omega^{0}_{\text{TC}}/(2 \pi) = 7.445~\rm{GHz}$.
The control pulses are designed in such a way that $\delta \omega_{\text{TC}}$ is 0 at the beginning and the end of the simulation.
The model can be accurately described using the the first two states of the qubits and of the TC since the higher energy states do not affect the process of interest, namely the population transfer between the states $\ket{0}$ and $\ket{1}$ of Q1 and Q2.
The eigenvectors of the drift Hamiltonian in Eq.~\eqref{eq3}, labeled $\ket{q_{1}q_{2}q_{\text{TC}}}$, are used to identify the $2^3$ system states, see Fig.~\ref{fig1}(b).
Due to the modest size of the problem, we do not need to introduce projectors as described in Eq.~\eqref{eq_proj}.
When $\delta\omega_\text{TC}$ is swept from $0$ to $-3~\rm{GHz}$, we observe two avoided level crossings between the TC state and the qubit states, see Fig.~\ref{fig1}(c). 
The associated nonadiabatic coupling terms, obtained with the Hellmann-Feynman expression~\footnote{The nonadiabatic coupling terms $d_{jk}$ between pairs of the full system \eqref{eq1} eigenstates $\ket{\psi_{j}}$ and $\ket{\psi_{k}}$ with corresponding eigenenergies $\varepsilon_{j}$ and $\varepsilon_{k}$ given by relation
$
\hat{H}(\delta \omega_{\text{TC}}) \ket{\psi_{j}(\delta \omega_{\text{TC}})} = \varepsilon_{j}(\delta \omega_{\text{TC}}) \ket{\psi_{j}(\delta \omega_{\text{TC}})}
$
for a certain $\delta \omega_{\text{TC}}$ pulse value are determined with the Hellmann-Feynman expression \cite{Singh1989}
$
d_{jk}(\delta \omega_{\text{TC}}) = \frac{\bra{\psi_{j}(\delta \omega_{\text{TC}})} \frac{\partial \hat{H}(\delta \omega_{\text{TC}})}{ \partial \delta \omega_{\text{TC}}} \ket{\psi_{k}(\delta \omega_{\text{TC}})}}{\varepsilon_{j}(\delta \omega_{\text{TC}}) - \varepsilon_{k}(\delta \omega_{\text{TC}})}
$
}
are shown in Fig.~\ref{fig1}(d). 
The LCT algorithm will make use of these avoided level crossings to transfer population between the two qubits.

\begin{center}
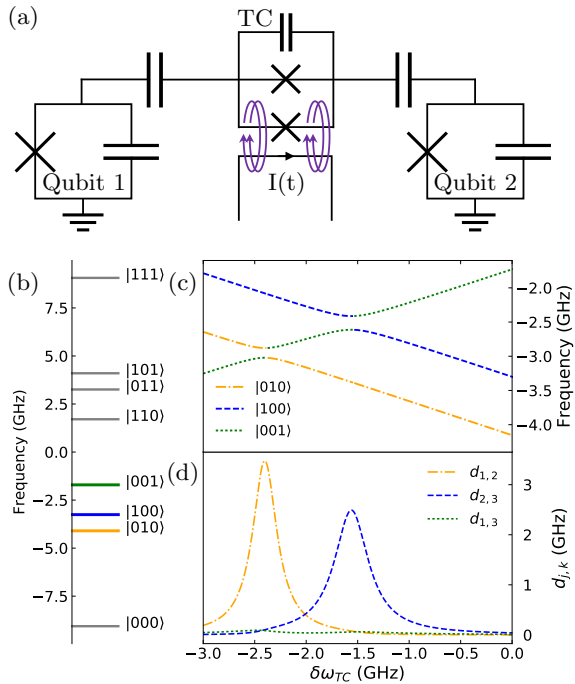
\begin{figure}[t]
\begin{tikzpicture}
\node at (0,4.25) {\includegraphics[width=0.48\textwidth, clip, trim=100 100 475 225]{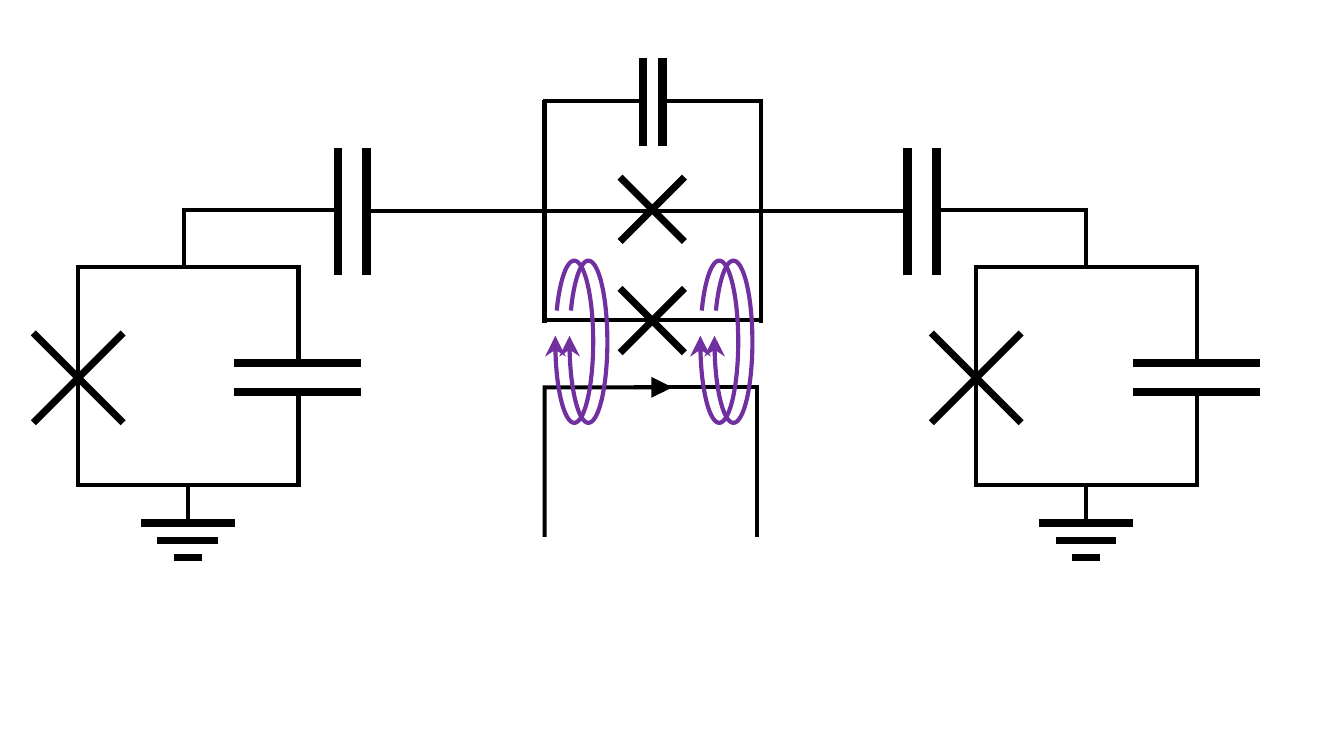}};
\node at (0,0) {\includegraphics[scale=0.4]{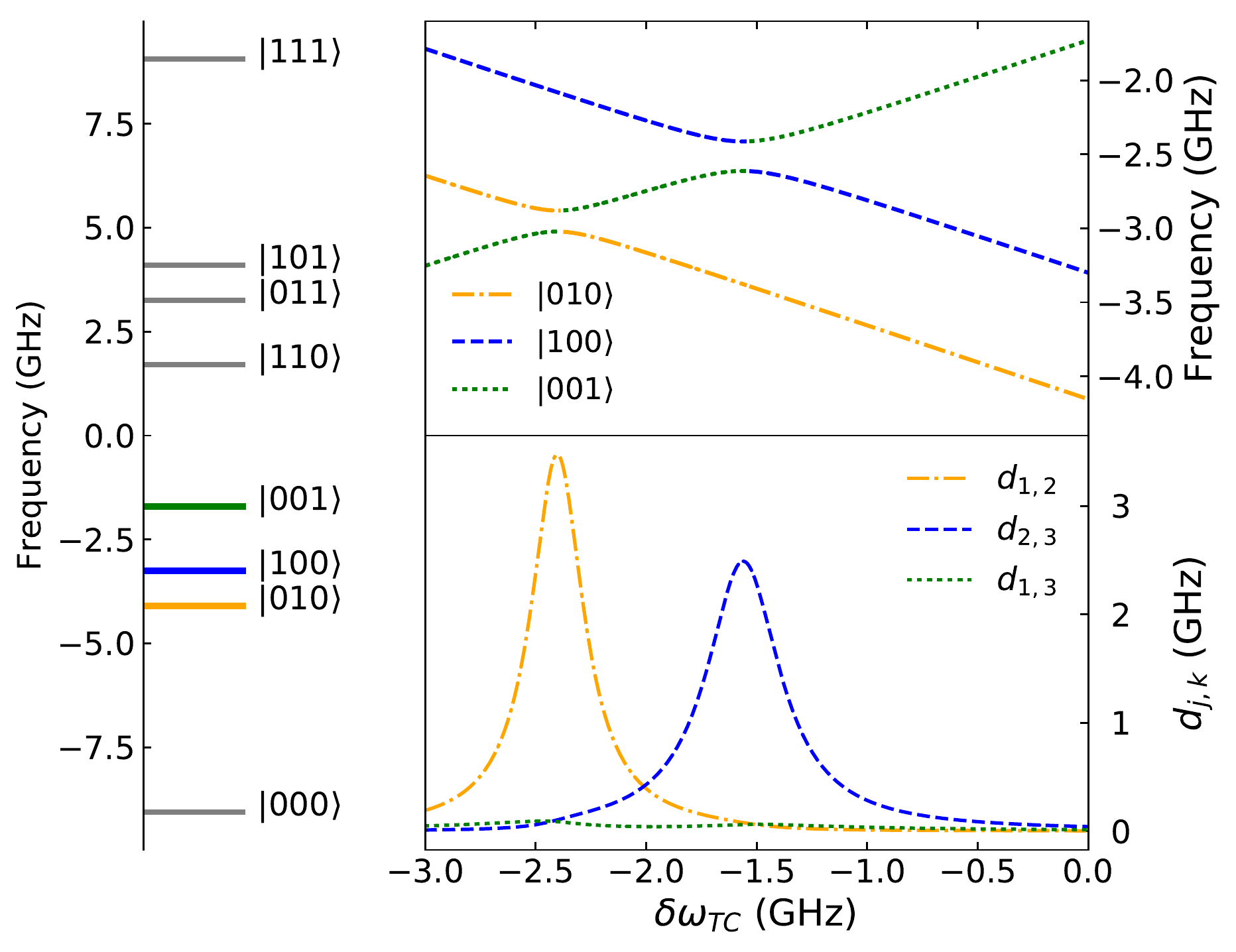}};
\node at (-3.55,6.00) {(a)};
\node at (-1.45,2.475) {(c)};
\node at (-1.45,-0.05) {(d)};
\node at (-3.55,2.475) {(b)};
\node at (-0.08,3.80) {I(t)};
\node at (-2.75,3.80) {Qubit 1};
\node at ( 2.40,3.80) {Qubit 2};
\node at (-0.50,6.00) {TC};
\end{tikzpicture}
\caption{(a) Sketch of two fixed-frequency transmon qubits coupled using a tunable coupler. 
(b) Energy level scheme of the undriven system.
(c) Evolution of the eigenstates of the full Hamiltonian in Eq.~(\ref{eq1}) as a function of $\delta \omega_{\text{TC}}$. The labels $\ket{010}$ (dotted-dashed orange), $\ket{100}$ (dashed blue) and $\ket{001}$ (dotted green) refer to the eigenstates of the Hamiltonian when the coupling is set to zero.
(d) The nonadiabatic couplings $d_{1,2}$ (dotted-dashed orange), $d_{2,3}$ (dashed blue) and $d_{1,3}$ (dotted green), as functions of $\delta \omega_{\text{TC}}$.
}
\label{fig1}
\end{figure}
\end{center}

\section{Results and Discussion}
\label{sec: results_and_discussion}

\subsection{LCT pulse}
\label{subsec:bare_lct}

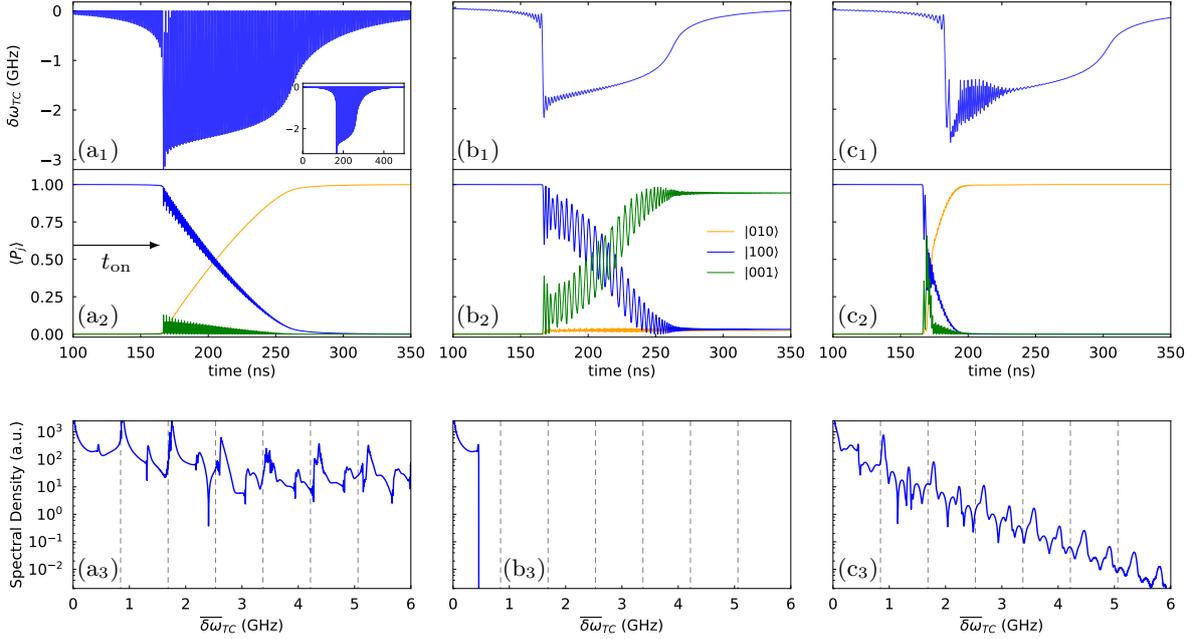
\begin{figure*}[htbp!]
\centering
\begin{tikzpicture}
\node at (0,0) {\includegraphics[width=0.98\textwidth]{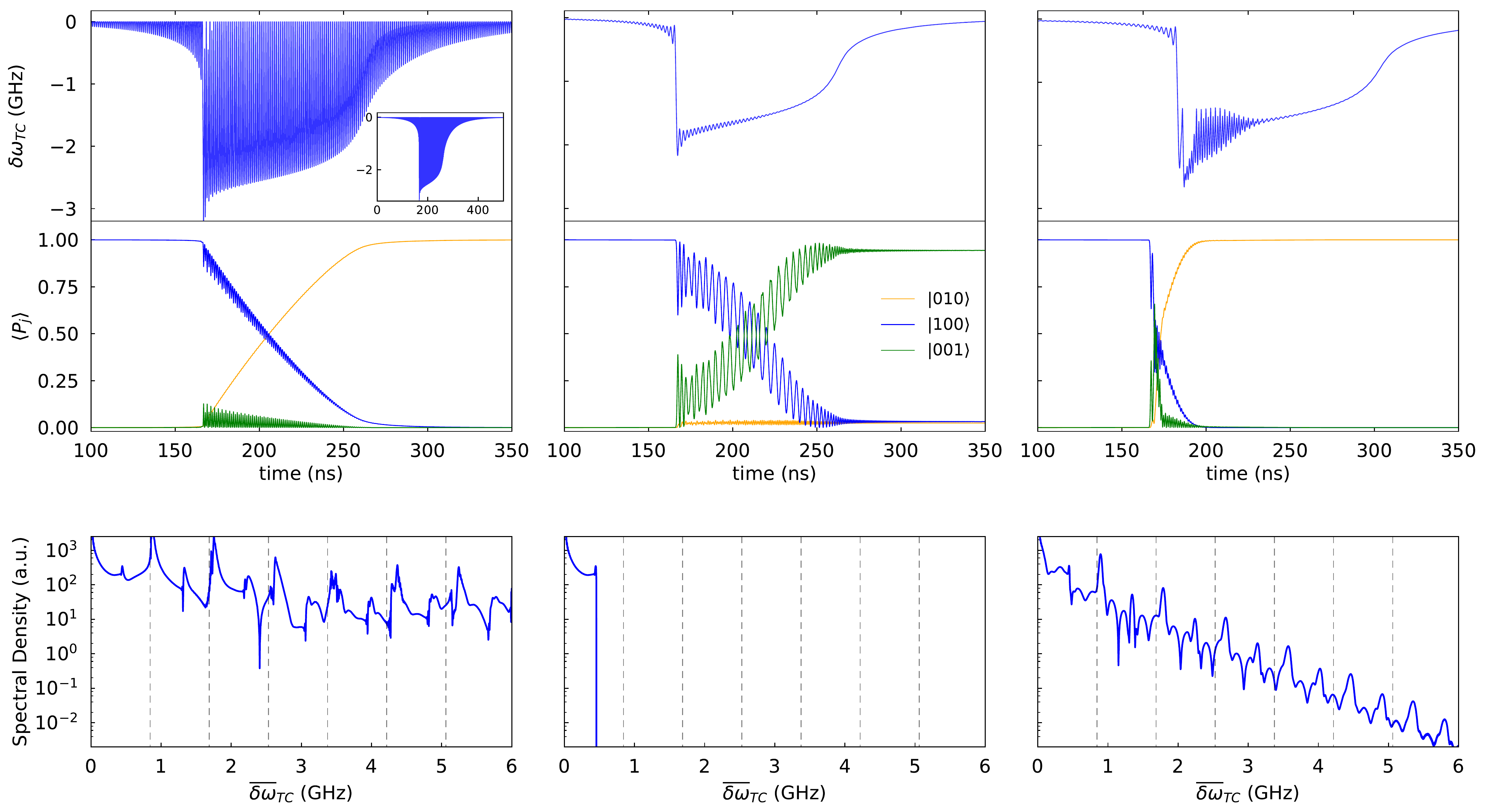}};
\draw[-latex] (-6.88,0.95) -- +(1.15,0.0) node[pos=0.5,below] {$t_\text{on}$};
\node at (-6.55,2.2) {($\text{a}_1$)};
\node at (-6.55,0) {($\text{a}_2$)};
\node at (-6.55,-3.35) {($\text{a}_3$)};
\node at (-1.55,2.2) {($\text{b}_1$)};
\node at (-1.55,0) {($\text{b}_2$)};
\node at (-0.94,-3.35) {($\text{b}_3$)};
\node at (3.45,2.2) {($\text{c}_1$)};
\node at (3.45,0) {($\text{c}_2$)};
\node at (3.45,-3.35) {($\text{c}_3$)};
\end{tikzpicture}
\caption{
($\text{a}_1$) LCT pulse designed to transfer population from state $\ket{100}$ to state $\ket{010}$ (Inset: full 450 ns pulse). 
The parameter $\lambda$ was set to
$12500$.
($\text{b}_1$) Frequency filtered pulse $\delta\omega_\text{TC}^\text{filt}(t)$ used as an initial condition to design the second local control pulse. 
($\text{c}_1$) LCT pulse designed to transfer population from state $\ket{100}$ to state $\ket{010}$ when using the pulse in ($\text{b}_1$) as an initial condition. 
($\text{a}_2$), ($\text{b}_2$) and ($\text{c}_2$) population transfer resulting from the pulses in ($\text{a}_1$), ($\text{b}_1$) and ($\text{c}_1$), respectively. 
($\text{a}_3$), ($\text{b}_3$) and ($\text{c}_3$) Fourier transforms of the pulses in ($\text{a}_1$), ($\text{b}_1$) and ($\text{c}_1$), respectively. 
The dashed lines indicate the harmonics corresponding to frequency differences between the qubits.
}
\label{FigLC1}
\end{figure*}

In this section, we design a LCT 
pulse that achieves population transfer from the state $\ket{100}$ to the state $\ket{010}$, i.e. that brings the excitation from Q1 to Q2. 
We assume that the TC is biased at the flux sweet spot $\Phi(t=0)=0$.
Since the initial and final states are orthonormal, we use the state preparation in Eq.\ (\ref{Eqn:seed}) with $\eta=10^{-6}$ to initialize the LCT algorithm.

Figure \ref{FigLC1}($\text{a}_1$) shows a $150~\rm{ns}$ long LCT pulse obtained for $\lambda=12500$. 
This pulse makes the tunable coupler energy level oscillate between the two avoided level-crossings depicted in Fig.~\ref{fig1}(c). 
As the TC $\ket{001}$ state passes through the first avoided crossing at $-1.56~\rm{GHz}$ a fraction of the qubit population in $\ket{100}$ is transferred to the TC. 
Part of this population is then transferred to the second qubit (state $\ket{010}$) once the second avoided crossing at $-2.40~\rm{GHz}$ is reached. 
The TC oscillates with a complicated frequency pattern dominated by the harmonics of the transition between the two qubits, $(\omega_1-\omega_2)/(2\pi)=859~\rm{MHz}$ and by other components below $1~\rm{GHz}$ as shown by the power spectrum of the pulse in Fig.~\ref{FigLC1}($\text{a}_3$). 
It is important to note that, despite the many frequencies appearing on the Fourier transform of the LCT pulse, no other transition further than the targeted ones are excited during LCT process. 
As expected from Eqs. \eqref{eq7} and \eqref{eq8}, after an initial delay $t_\text{on}$ of about $170~\rm{ns}$, the population of the target state ($\ket{010}$) increases monotonically with time while the populations of the other states considered in the simulation shows important high frequency oscillations. 
At the end of the transfer process ($\sim 300~\rm{ns}$), the initial population has been almost entirely transferred to the target state, achieving a mismatch $1-P_{\ket{010}}$ of less than  $10^{-6}$, where $P_{\ket{010}}$ is the population of the target state.
While very promising, this first `high fidelity' LCT pulse has a highly complex spectrum and its implementation requires instruments with a large bandwidth.

\subsection{Optimization of LCT pulses}
\label{subsec:lct_optimization}

Because of limits set by the control instruments, large bandwidth pulses are impractical to generate. 
We therefore need a procedure to refine the LCT pulse, which allows to confine the bandwidth within a reasonable range.
To this end, we apply a high frequency filter 
to the LCT pulse obtained in the previous section and use it as a `reference' to generate an improved pulse using the LCT algorithm. 
This new reference corresponds to the term $\delta\omega^\text{filt}_\text{TC}(t) $ in Eq.~\eqref{Eq_LCT_in_3_parts}.
In practice, we decompose the new LCT pulse into three different components
\begin{align} \label{Eq_LCT_in_3_parts}
\omega_\text{TC}(t)=
\omega^0_\text{TC} +  \delta\omega^\text{filt}_\text{TC}(t) + \delta\omega^{\text{lct},2}_\text{TC}(t) \, .
\end{align}
Only the component $\delta\omega^{\text{lct},2}_\text{TC}(t)$, initially set to $0$, will be generated on-the-fly using the LCT algorithm,
while the first two terms 
are kept fixed.
As in the previous section, the pulse $\delta\omega_\text{TC}^{\text{lct},2}$ is shaped on-the-fly using the requirement that the right-hand side of Eq.~\eqref{eq7}, $\partial_t \langle \hat{P}_{j} \rangle$, remains positive.

The filtered pulse $\delta\omega^\text{filt}_\text{TC}(t)$ in Fig.~\ref{FigLC1}($\text{b}_1$) is obtained by applying a high frequency cut-off at $0.4~\rm{GHz}$ to the pulse in Fig.~\ref{FigLC1}($\text{a}_1$). The corresponding spectra before and after the application of the filter are shown in Fig.~\ref{FigLC1}($\text{a}_3$) and Fig.~\ref{FigLC1}($\text{b}_3$), respectively. 
This operation removes much of the complex structure of the pulse while preserving its overall shape (Fig.~\ref{FigLC1}($\text{b}_1$)).
As expected, the pulse composed by the first two components in Eq.~\eqref{Eq_LCT_in_3_parts} fails to transfer the population to the target qubit (Fig.~\ref{FigLC1}($\text{b}_2$)). 
However, using LCT we can generate a new time-dependent field, i.e. $\delta\omega^{\text{lct},2}_\text{TC}(t)$ in Eq.~\eqref{Eq_LCT_in_3_parts} with coupling parameter $\lambda_2$, which restores this property.
In particular, we are able to design new LCT pulses with a narrow bandwidth and an error $1-P_{\ket{010}}<10^{-6}$ using a wide range of $\lambda_2$ in the interval $[100, 1000]$, see Fig.~\ref{FigLC1}($\text{c}_1$-$\text{c}_3$).
In addition, the population transfer is now completed in only $\sim 30~\rm{ns}$ (see Fig.~\ref{FigLC1}($\text{c}_2$)) compared to the initial $120~\rm{ns}$ obtained with the first LCT run described in Sec.~\ref{subsec:bare_lct} (Fig.~\ref{FigLC1}($\text{a}_2$)).
This improvement results from the nature of the `reference' pulse, i.e. the sum $\omega_\text{TC}^{0} + \delta\omega^\text{filt}_\text{TC}(t) $, forcing the TC frequency in the energy range that matches the separation between the two avoided crossings shown in Fig.~\ref{fig1}(c). 
Note that the truncation of the power spectrum in Fig.~\ref{FigLC1}($\text{c}_3$) above 1GHz (1.5GHz) without further optimization will reduce the fidelity to $10^{-4}$ ($10^{-5}$).

\subsection{Reverse processes}
\label{subsec:reverse}

So far, the LCT pulses were generated to accomplish a well defined transition from a given initial state to a final state.
Therefore, we cannot expect that by applying the same pulse to the final state it can revert the process and transfer the population back to the initial state.  
For instance, applying the pulse in Fig.~\ref{FigLC1}($\text{a}_1$) (generated for the population transfer from $\ket{100}$ to $\ket{010}$) to the reverse process (from $\ket{010}$ back to $\ket{100}$) we only achieve an imperfect transfer that leaves $29\%$ of the population in the TC.
Interestingly, we found that the amount of population trapped in the TC is particularly sensitive to the value of the parameter $\lambda_2$.
Therefore, the LCT pulse can be further optimized to increase the efficiency of the reverse transfer by tuning  $\lambda_2$. 
Note that changes to $\lambda_2$ do not affect the success of the population transfer from $\ket{100}$ to $\ket{010}$, since the conditions (initial and final states) and the reference pulse $\delta\omega_\text{TC}(t)$ are kept fixed. 

Exploiting this fact, we illustrate a procedure for the recursive optimization of the direct and reverse population transfers between the states $\ket{100}$ and $\ket{010}$, starting from the bandwidth optimized pulse derived in Sec.~\ref{subsec:lct_optimization}.
For an initial choice of  $\lambda_2$, we derive a first LCT pulse for the direct process ($\ket{100}$ to $\ket{010}$) and then test it for the reverse transfer  ($\ket{010}$ to $\ket{100}$). 
If this fails to accomplish a population transfer back to the initial state $\ket{100}$ with an error $1-P_{\ket{100}}$ less than $10^{-6}$ we update the parameter $\lambda_2$ and recompute the pulse using the LCT algorithm.
This procedure is repeated until the reverse population transfer fidelity reaches a maximum.
The Nelder-Mead algorithm \cite{Nelder1965} is used to optimize $\lambda_{2}$.
In some cases, we noticed that maximizing the population transferred during the reverse process required a change of the frequency cut-off values for $\delta\omega_\text{TC}^\text{filt}(t)$.
\begin{figure}[t]
\centering
\begin{tikzpicture}
\node at (0,0) {\includegraphics[width=0.45\textwidth]{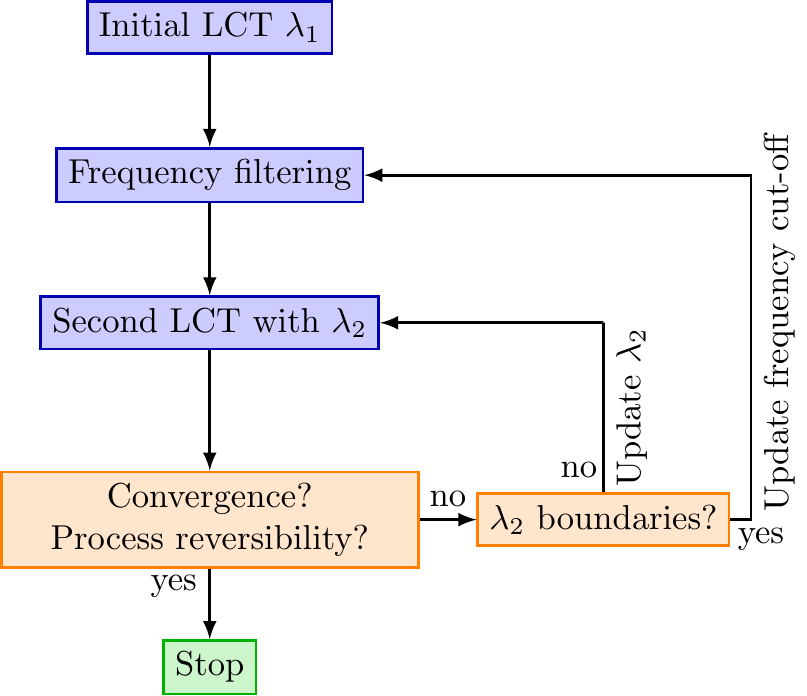}};
\node at (-2.1cm,2.40cm) {ref. pulse: $\omega^0_{\text{TC}}$};
\node at (-1.35cm,-0.25cm) {ref. pulse: $\omega^0_{\text{TC}}+\delta\omega^{\text{filt}}_{\text{TC}}(t)$};
\end{tikzpicture}
\caption{
Flow chart showing the iterative procedure used to obtain pulses with a smaller bandwidth and capable of transferring population when the initial and target states are exchanged. 
A new LCT calculation is performed each time the parameter $\lambda_2$ is updated or when a new frequency cut-off is applied.
}
\label{Scheme1}
\end{figure}
A flow chart of the algorithm used to obtain a narrow bandwidth pulse able to transfer the qubit population in both directions is shown in Fig.~\ref{Scheme1}.   
For the setup in Fig.~\ref{fig1} and the parameter discussed in Section~\ref{subsec: system}, the produced LCT pulse is given in Fig.~\ref{FigLC4}(a) together with the population dynamics for the direct and reverse processes respectively shown in Fig.~\ref{FigLC4}(b) and (c). 
The final LCT pulse in Fig.~\ref{FigLC4}(a) can further be used to inspire a new class of fully analytical and ultrashort pulses that can be used for state preparation (see Appendix~\ref{Appendix1}).

\begin{figure}[t]
\centering
\begin{tikzpicture}
\node at (0,0) {\includegraphics[width=0.475\textwidth]{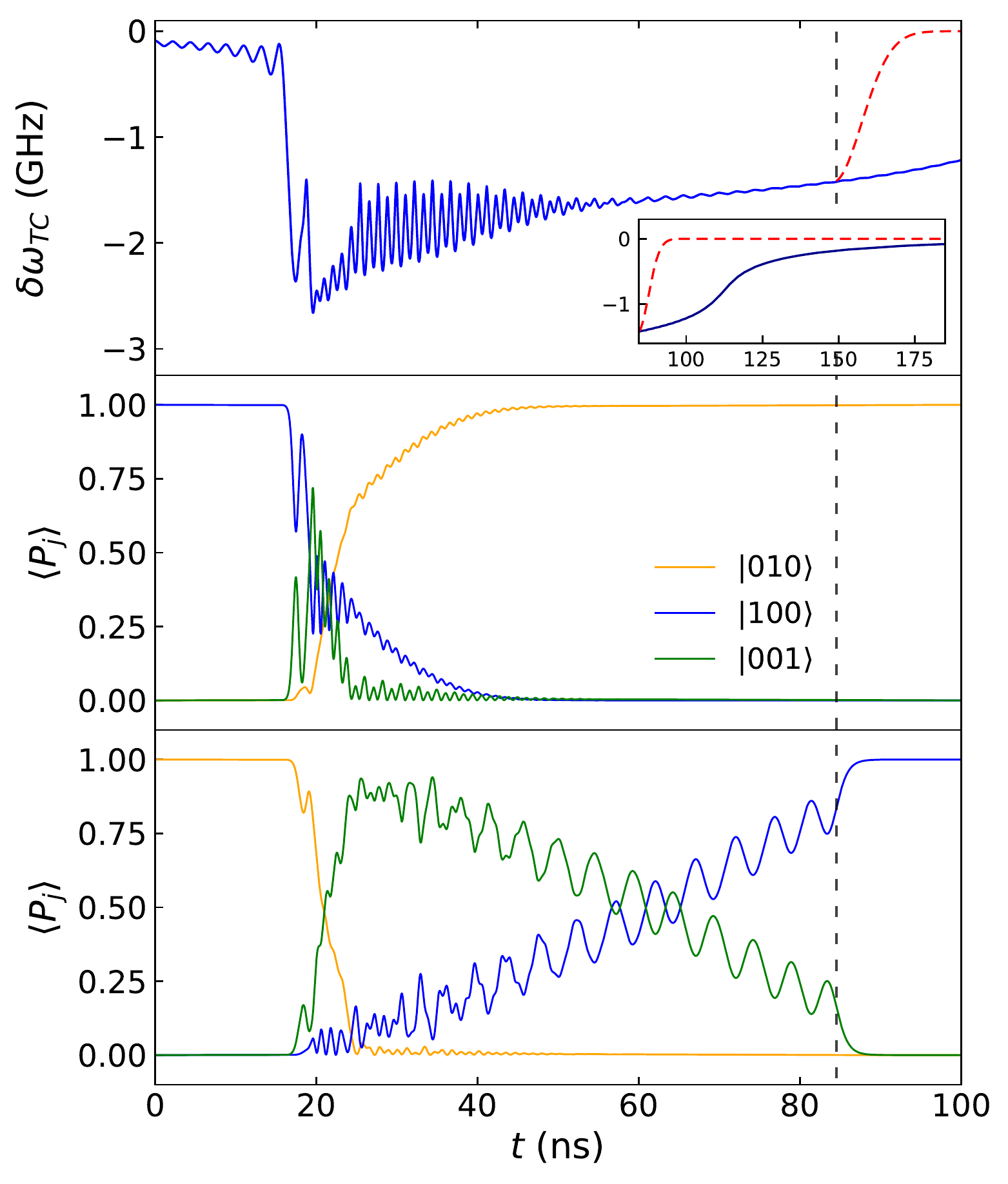}};
\node at (2.5,4.5) {$\tau$};
\node at (-2.3,1.9) {(a)};
\node at (-2.3,-0.55) {(b)};
\node at (-2.3,-3.15) {(c)};
\end{tikzpicture}
\caption{
(a) Final pulse generated by the algorithm depicted in Fig.~\ref{Scheme1} (final frequency cut-off at $0.45~\rm{GHz}$, $\lambda_{2}=437.4$). 
The inset shows the tail of the pulse (after the time $\tau$), which can be substituted with the half-Gaussian function (dashed red line). 
(b) Evolution of the system when all the population is initially in Q1.
(c) Evolution of the system when all the population is initially in Q2. 
The populations shown in (b) and (c) are calculated using the pulse with the shortened tail.
}
\label{FigLC4}
\end{figure}
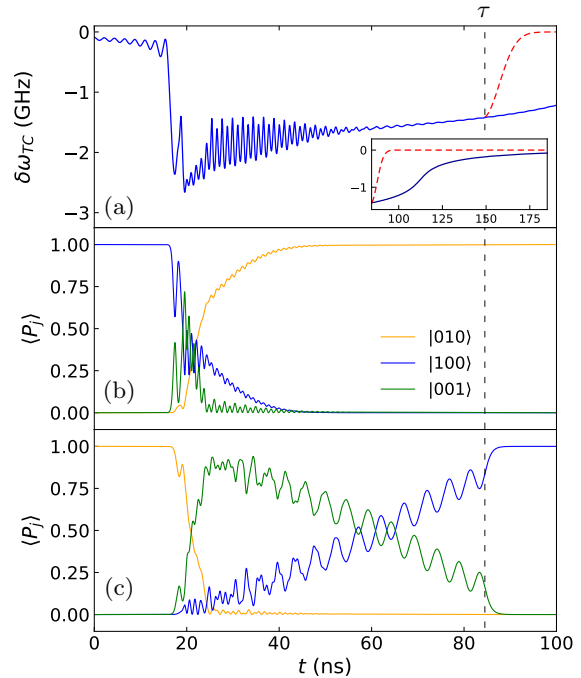

\subsection{Pulse truncation}
\label{subsec:truncation}

The pulses obtained using the algorithm shown in Fig.~\ref{Scheme1} still have a long tail in the time domain that is inherited from the original, fixed $\lambda$, LCT calculation (Section~\ref{subsec:bare_lct}).
Since the tail does not contribute to the population transfer, see e.g. Fig.~\ref{FigLC1}($\text{c}_2$), the pulses can be shortened by imposing a Gaussian decay after a critical time $\tau$ using the half Gaussian function $\alpha \exp\{-(t-\tau)^2/(2\sigma^{2})\}$ for $t \geq \tau$ shown in Fig.~\ref{FigLC4}(a).
The optimal value of $\tau$ is obtained by including it in the optimization process shown in Fig.~\ref{Scheme1}, while its
initial value is selected as the time required by the original pulse in Fig.~\ref{FigLC4}(a) to reach 99\% of population transfer for the reverse process in Fig.~\ref{FigLC4}(c).
For a chosen $\sigma$ value, this leads to an optimized pulse where population transfer fidelities $1-P_{\ket{010}}$ and $1-P_{\ket{100}}$ are both less than $10^{-6}$ for the forth and back population transfer, respectively.
Finally, the choice of the parameter $\alpha$ is imposed by the need to guarantee continuity at the transition point.
 
\section{Conclusion}
\label{sec:conclusion}
In this work we propose using local control theory (LCT) to manipulate qubit populations in an architecture where fixed-frequency superconducting qubits are coupled using tunable couplers.
Given the initial and target states, LCT constructs a pulse on-the-fly by computing the time evolution 
only once.
The only tunable parameter is the intensity of the applied pulse (controlled by $\lambda$ in Eq.~\eqref{eq8}).
$\lambda$ influences the shape and length of the resulting LCT pulse, giving the possibility to shorten the transfer time below 50 ns while keeping a high fidelity for the process.

The LCT algorithm was extended to design pulses that can achieve a complete population transfer in both directions between the initial and the target states.
This extension of the LCT algorithm comprises an additional optimization step over the parameter $\lambda$.
LCT can also serve as a starting point for a deterministic procedure to further reduce the complexity of the pulse, see Appendix.
This opens up a new avenue of research to design efficient gates for different applications of quantum computing~\cite{moll_quantum_optimization_2018,Egger2018, Barkoutsos2018a}. 

Further work will investigate the sensitivity of LCT pulses to the different parameters characterizing the model Hamiltonian (Eq.~\eqref{eq1}), as well as using LCT in systems with more elements where frequency crowding may become an issue. 
Such systems could for instance include $n>2$ qubits coupled to the same tunable coupler.

\section{Acknowledgment}

The authors acknowledge stimulating discussions with Marco Roth and Nikolaj Moll. We also acknowledge generous computational time from the Croatian National Grid Infrastructure (CRO-NGI) and the Irish Centre for High-End Computing (ICHEC). 
This work has received funding from the European Union's Horizon 2020 research and innovation program under the grant agreement No. 676531 (project E-CAM).

\bibliography{LCT_references}

\appendix

\section{Analytic state preparation pulse}
\label{Appendix1}

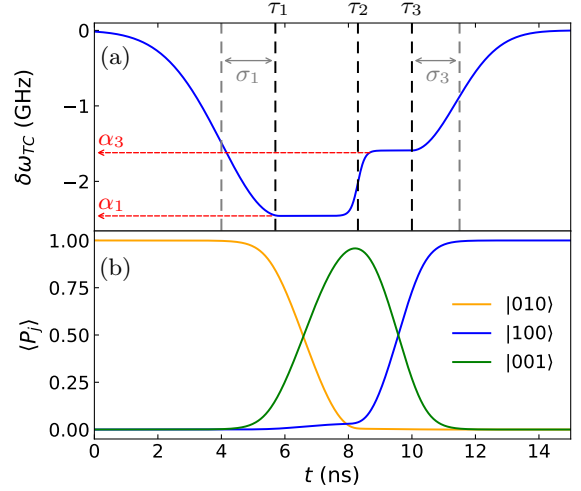
\begin{figure}[t]
\centering
\begin{tikzpicture}
\node at (0,0) {\includegraphics[width=0.475\textwidth]{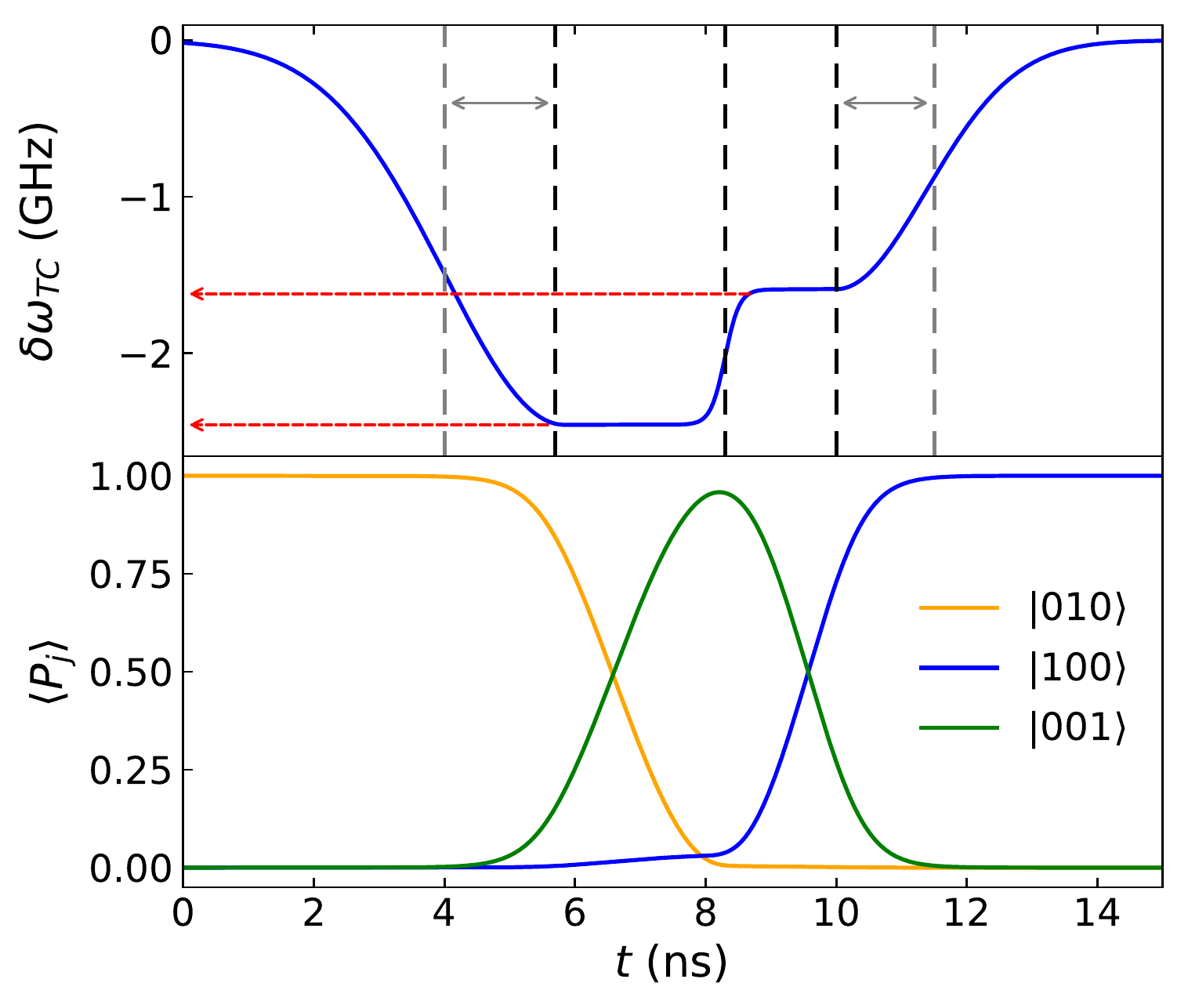}};
\node at (-2.35,2.6) {(a)};
\node at (-0.2,3.25) {\textcolor{black}{$\tau_{1}$}};
\node at (0.85,3.25) {\textcolor{black}{$\tau_{2}$}};
\node at (1.55,3.25) {\textcolor{black}{$\tau_{3}$}};
\node at (-0.6,2.35) {\textcolor{gray}{$\sigma_{1}$}};
\node at (1.9,2.35) {\textcolor{gray}{$\sigma_{3}$}};
\node at (-2.4,1.5) {\textcolor{red}{$\alpha_{3}$}};
\node at (-2.4,0.65) {\textcolor{red}{$\alpha_{1}$}};
\node at (-2.35,-0.2) {(b)};
\end{tikzpicture}
\caption{
(a) Time optimal pulse shape obtained from Eq.~\eqref{eq10} accomplishing the $\ket{010} \rightarrow \ket{100}$ transfer using the TC $\ket{001}$.
Optimal parameter values: $\alpha_{1}=-2.457~\rm{GHz}$, $\alpha_{3}=-1.591~\rm{GHz}$, $\tau_{1}=5.8~\rm{ns}$, $\tau_{2}=8.3~\rm{ns}$, $\tau_{3}=10.0~\rm{ns}$, $\sigma_{1}=1.83~\rm{ns}$, $\sigma_{2}=0.2~\rm{ns}$, $\sigma_{3}=1.37~\rm{ns}$.
(b) Evolution of the populations during the pulse shown in (a). 
}
\label{FigLC6}
\end{figure}

Exploiting the fact that population transfer between states are fastest in regions of avoided crossings we construct an analytic pulse which drives the system into regions of largest nonadiabatic couplings in order to further reduce the population transfer times.
Based on the pulse in Fig.~\ref{FigLC4}(a), we construct the following analytical pulse 
\begin{equation}
\delta \omega_{\text{TC}}(t) = \left\{ \begin{array}{ll} \alpha_{1} \exp(-\frac{1}{2}(\frac{t-\tau_{1}}{\sigma_{1}})^{2}) &;\, t < \tau_{1} \\ 
& \\
\frac{1}{2}(\alpha_{3} + \alpha_{1}) + \frac{1}{2}(\alpha_{3} \\ - \alpha_{1}) \tanh(\frac{t-\tau_{2}}{\sigma_{2}})  &; \, \tau_{1} \leq t \leq \tau_{3} \\
& \\
\alpha_{3} \exp(-\frac{1}{2}(\frac{t-\tau_{3}}{\sigma_{3}})^{2}) &; \, t > \tau_{3} \end{array} \right.~\label{eq10}
\end{equation} 
with the aim of minimizing the duration of the state preparation pulse.
Here $\alpha_{i}$ are the amplitudes and $\sigma_{1}$ and $\sigma_{3}$ the decay times of two  half-Gaussian pulse envelopes connected by a switching function with a slope controlled by $\sigma_{2}$, see Fig.~\ref{FigLC6}(a).
The values of these parameters were obtained using an optimization-with-bounds procedure from the sequential least square programming algorithm \cite{kraft1988}, which enforces a complete population transfer from $\ket{010}$ to $\ket{100}$. 
For the initial conditions, the values of $\alpha_{1}$ and $\alpha_{3}$ were set equal to the energies of the second and first avoided crossings, respectively, while $\tau_{2}-\tau_{1}$ and $\tau_{3}-\tau_{1}$ were set to the population transfer times obtained from LCT for the  $\ket{010} \rightarrow \ket{001}$ and $\ket{100} \rightarrow \ket{010}$, respectively 
(see Fig.~\ref{FigLC4}(c) and (b), respectively).
In our simulations, $\tau_{1}$ is chosen arbitrarily in the range between 5 to 6 ns.
Initial $\sigma_{i}$ parameters were set close to zero and relaxed during a second optimization step once the $\alpha$-s and $\tau$-s were fully determined. 
This leads to the generation of the smooth final pulse shown in Fig.~\ref{FigLC6}(b).
This analytical pulse shape results in a sequential population transfer from Q2 to TC and finally to Q1 with a total fidelity $1-P_{\ket{100}} < 10^{-6}$.
The reverse population transfer (from Q1 to Q2 via the TC) is achieved by inverting the pulse in the time domain.
This pulse is similar to the case where frequency tunable elements are used to shuttle population to and from different elements in a larger quantum system \cite{Mariantoni2011_photonshell}.
Interestingly, the final pulse duration is short ($\sim15~\rm{ns}$) compared to the coherence times in state-of-the-art experiments.

\end{document}